# Displacement current: examples that go beyond the beaten path

**Álvaro Suárez[1], Martín Monteiro[2], Arturo C. Martí[3]**

1. Consejo de Formación en Educación, Instituto de Profesores Artigas, Montevideo, Uruguay
alsua@outlook.com
2. Universidad ORT Uruguay, Montevideo, Uruguay
monteiro@ort.edu.uy
3. Instituto de Física, Facultad de Ciencias, Udelar, Montevideo, Uruguay
marti@fisica.edu.uy

**Abstract**

The Ampere-Maxwell's law and the displacement current constitute one of the most difficult aspects of electromagnetic theory for students in introductory electromagnetics courses. Here we present a set of examples that go beyond the classical ones usually discussed in introductory textbooks. In-depth analysis of these examples allows to develop a deeper understanding of electromagnetic theory even in students who have not acquired the full mathematical toolkit of the more advanced courses.

## I. Introduction

Electromagnetism is prominent in university science and engineering courses worldwide. The vast majority of general physics courses follow the historical development of electromagnetic theory starting with static phenomena and then adding the dynamical aspects of electromagnetism ([1, 2] and references therein). A key moment in the development of this theory is the introduction of the displacement current, which led to the generalization of Ampère's law, now known as Ampère-Maxwell (AM) law. This advance provided a unified view of electromagnetism and paved the way for the study of electromagnetic waves, linking their generation and propagation to light and optical phenomena.

The historical development of displacement current is closely related to the development of classical electromagnetic theory, a process that spanned much of the 19th century. James Clerk Maxwell's contributions, deeply influenced by Michael Faraday's field concepts, were instrumental in the shift from a mechanical ether model to a field-based perspective [3]. In his 1861 work "On Physical Lines of Force", Maxwell introduced the displacement current by establishing a connection between electrical conduction in conductors and electrical displacement in insulators. This step allowed Maxwell to develop a continuity equation similar to the one used today and to generalize Ampère's law by including the displacement current as a necessary term to satisfy the continuity equation [4]. In 1864, Maxwell published "A Dynamical Theory of the Electromagnetic Field", where he further clarified the concept of displacement current, identifying it as a distinct type of current that contributes to the total current along with the conduction current [5]. Finally, in his 1873 work "A Treatise on Electricity and Magnetism", Maxwell emphasized that the inclusion of displacement current was essential to reconcile the laws of electromagnetism with the existence of open conduction currents [4].

Despite its foundational importance, the displacement current typically receives only brief treatment in comparison with other fundamental phenomena and features in significantly fewer textbook problems [6, 7]. For this reason, we perceive this subject as having a mysterious halo for students. Why does it appear? Can it be measured? Does it generate a magnetic field? [8, 9, 10]. Its intrinsic conceptual difficulty and limited treatment could be at the heart of some conceptual student difficulties described in the literature. In particular, it has been reported that students have difficulty recognizing the presence of a displacement current and that a time-varying electric field is necessarily linked to a magnetic field [11]. In addition, interferences have been observed between the concepts involved in Faraday's and AM laws, confusions between the concepts of flux and circulation, and a difficulty in recognizing the concept of displacement current apart from the 'standard' example related to the charge of a capacitor [11].

Following the usual notation we denote the electric field vector with **E**, the magnetic field as **B**, $\varepsilon_0$ is the vacuum permittivity and $\mu_0$ the vacuum permeability of free space. The integral expression of the AM law can be written as

$$\oint \mathbf{B} \cdot \mathbf{dl} = \mu_0 I_C + \mu_0 I_D \qquad (1)$$

where $I_C$ is the conduction current related to the movement of electric charges. The second term in the r.h.s. is due to the presence of time-varying electric fields, however, for historical reasons it is usually referred to as the displacement current

$$I_D = \varepsilon_0 \frac{d}{dt} \int \mathbf{E} \cdot \mathbf{dA} \qquad (2)$$

In this paper we analyze the role of the displacement current and the Ampère-Maxwell's law and show that a more in-depth treatment provides the opportunity to explore central issues in electromagnetism. We discuss in detail some of the examples commonly found in textbooks and also propose novel situations and exercises specifically designed for introductory courses in electromagnetism, taking into consideration the learning difficulties that have been previously reported [11, 12, 13].

## 2. The sources of the electromagnetic field: analyzing more in-depth the standard example

The most ubiquitous example for the treatment of the displacement current is the charging of a parallel circular plate capacitor. The consideration of a closed curve *C* and two limiting surfaces $S_1$ and $S_2$ as shown in Fig. 1 is useful to show the limits of applicability of Ampère's law and to introduce the displacement current. Indeed, the application of Ampère's law to these surfaces leads to contradictory results, which motivates the need to generalize it and include new terms.

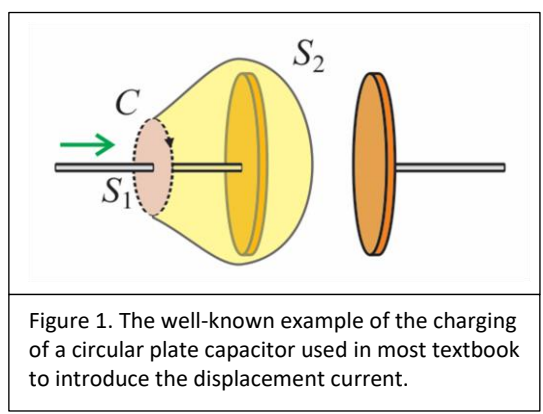

Figure 1. The well-known example of the charging of a circular plate capacitor used in most textbook to introduce the displacement current.

This example is also useful for discussing other aspects that normally do not receive much attention. The causality of electromagnetic fields is one of them. For this purpose, we apply the AM law, Eq. 1, using the curve C and alternately both surfaces. Considering $S_1$, crossed only by the conduction current, the magnetic field circulation results

$$\oint \mathbf{B} \cdot \mathbf{dl} = \mu_0 I_C \qquad (3)$$

One might rightly wonder why the displacement current due to the variable electric field inside the cable is not included in the integral of $S_1$. This is because we are considering almost ideal conductors, with almost zero resistivity, and since by Ohm's law, $E = \rho J$, then it turns out that inside the conductor $E \approx 0$.

Alternatively, since surface $S_2$ is crossed only by a displacement current the magnetic field circulation results

$$\oint \mathbf{B} \cdot \mathbf{dl} = \mu_0 I_D \qquad (4)$$

Thus, given the curve C, the magnetic field circulation is determined through the AM law and the same results should be obtained using both Eq. 3 or Eq. 4. This observation leads to contradictions with the causality view promoted in several textbooks [14, 15]. Indeed, using Eq. 3 we can observe that if we start from equation 3 we should conclude that the origin of the magnetic field lies in the conduction current while if we start from equation 4 the origin would be in the displacement current. However, both interpretations are incorrect, the AM law does not indicate causality relations but interdependence between quantities of the same physical entity: the electromagnetic field [10, 16].

Another surface bounded by the same curve, which can also be used to show that the AM law does not imply cause-and-effect relationships, is a cylindrical surface S3 with a radius smaller than that of the plates, as shown in Fig. 2. Let us note that the displacement current through this surface is less than on the $S_2$ in Fig. 1. Therefore, the only way for the magnetic field circulation to have the same value is the existence of conduction currents in the radial direction traversing the lateral surface of the cylinder [17]. These radial currents are responsible for increasing the charge of the capacitor. From this analysis, we conclude that depending on the surface we take, the magnetic field circulation can be determined by the conduction current (considering $S_1$), the displacement current ($S_2$) or both of them ($S_3$).

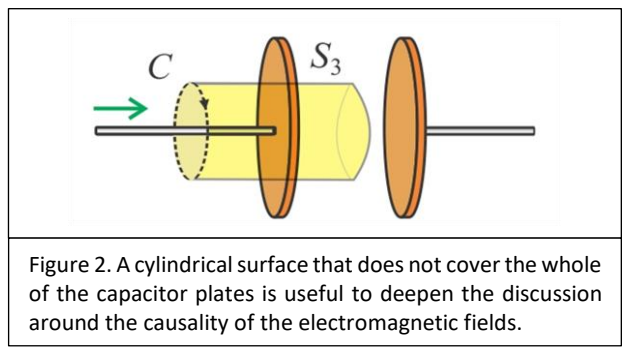

Figure 2. A cylindrical surface that does not cover the whole of the capacitor plates is useful to deepen the discussion around the causality of the electromagnetic fields.

The analysis of magnetic field circulation using different surfaces is useful for creating exercises that promote learning in depth in students. Exercises such as Ranking Tasks, Conflicting Contentions, or What, if anything, is Wrong Tasks? [18, 19] are ideal for challenging their ideas about the different features of the AM law. For example, given three surfaces, students could be asked to order the magnetic field circulation from largest to smallest, according to the surface used to calculate it, or to evaluate the validity of statements such as *The cause of the magnetic field at any point on the C curve depends on the surface you use to calculate it. If you take the $S_2$ surface, the cause will be the displacement current.*

To elucidate the origin of the magnetic field it is useful to incorporate into the discussion the conservation of charge and Gauss's law. If we consider the closed surface $S_0$, formed by the union of $S_1$ and $S_2$ the charge conservation can be expressed

$$I_C = \frac{dQ}{dt} \qquad (5)$$

where $Q$ is the total charge inside $S_0$, in this case, equal to the charge of the capacitor. The charge inside $S_0$ generates an electric field such that, according to Gauss's law, its flux through $S_0$ is equal to the charge inside, so Eq. 5 can be rewritten as

$$I_C = \frac{d}{dt}\int \varepsilon_0 \mathbf{E} \cdot \mathbf{dA} \qquad (6)$$

where we note that the r.h.s. is the displacement current $I_D$.

This analysis reinforces the concept that AM law does not establish cause-effect between the electric and magnetic fields but rather a link between quantities. Indeed, the displacement current $I_D$ in Eq. 4 is caused by the motion of charges accumulating on the capacitor plates and the magnetic field is originated in the conduction currents in the wire and in the capacitor plates [8, 20, 21].

## 3. The sources of the electromagnetic field: analyzing the magnetic field of a moving charge

The analysis of the magnetic field generated by a charge moving with linear rectilinear motion makes it possible to visualize the appearance of the displacement current in a situation outside the standard examples and also helps to clarify the cause-effect relationships in electromagnetic fields. Let us consider a charge $q$ moving along a straight line with constant velocity $v$, the magnetic field is usually described by the Biot-Savart's law

$$\mathbf{B} = \frac{\mu_0 q \mathbf{v} \times \mathbf{R}}{4\pi R^3} \qquad (7)$$

where $\mathbf{R}$ points from the charge to the point of interest.

To analyze the magnetic field from the AM law perspective we consider a surface $S$ and a circumference $C$ of radius $r$ that bounds it as shown in Fig. 3. As the charge approaches the curve, the electric field flux across the surface increases. Therefore, the non-zero magnetic field circulation along $C$ results

$$\oint \mathbf{B} \cdot \mathbf{dl} = \mu_0 \varepsilon_0 \frac{d}{dt}\int \mathbf{E} \cdot \mathbf{dA} \qquad (8)$$

and taking into account that the magnetic field is uniform along the curve we obtain

$$B = \frac{\mu_0 \varepsilon_0}{2\pi r} \frac{d}{dt}\int \mathbf{E} \cdot \mathbf{dA} \qquad (9)$$

It is possible to show that this expression leads to Biot-Savart's law for a moving charge if its velocity is much smaller than the speed of light [22]. This fact does not imply that the displacement current is the cause of the magnetic field generated by the moving charge. If so, the source would depend on whether we apply the Biot-Savart's or the AM law.

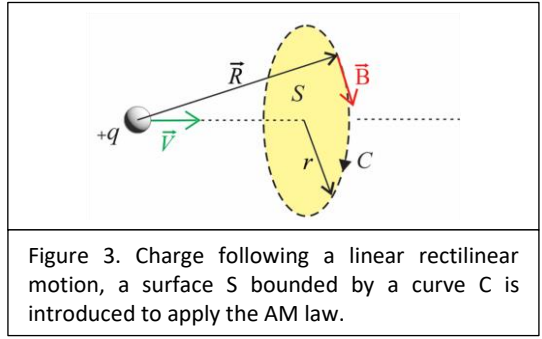

Figure 3. Charge following a linear rectilinear motion, a surface S bounded by a curve C is introduced to apply the AM law.

The situation analyzed could be used as an application of the AM law and the displacement current, as well as a starting point to analyze the limits of the validity of Ampere's law and to realize that it produces contradictory results in the presence of a time-dependent electric field. In this sense, we think that the characteristics of the problem make it ideal for developing a tutorial to introduce the AM law.

## 4. Charge conservation and AM law

Let us consider an initially charged spherical capacitor consisting of two concentric conducting shells filled with non-zero conductivity and electrical permittivity $\varepsilon$ material as shown in Fig. 4. The capacitor discharges through the dielectric with the charges moving radially. Given the spherical symmetry of these currents, the magnetic field vanishes, then, its circulation on any closed curve also vanishes [23]. One way to understand why the magnetic field is zero at any point P is by observing that the current density is uniform in all directions. This means that for every current element $dI$, there exists a symmetric counterpart $dI'$ with respect to P. Consequently, the magnetic field dB produced by $dI$ at P has the same magnitude but is opposite in direction to the magnetic field dB' produced by $dI'$. Therefore, $\mathbf{dB} + \mathbf{dB'} = 0$' for every pair of symmetric current

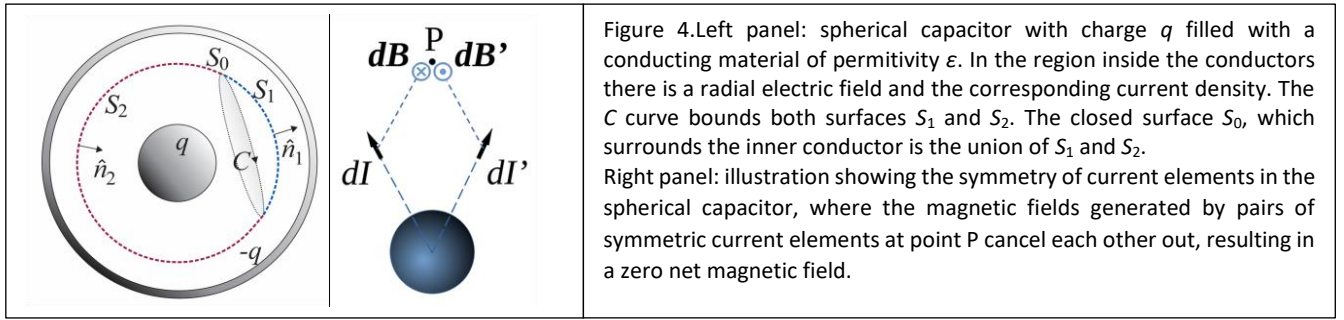

Figure 4. Left panel: spherical capacitor with charge $q$ filled with a conducting material of permitivity $\varepsilon$. In the region inside the conductors there is a radial electric field and the corresponding current density. The C curve bounds both surfaces $S_1$ and $S_2$. The closed surface $S_0$, which surrounds the inner conductor is the union of $S_1$ and $S_2$.
Right panel: illustration showing the symmetry of current elements in the spherical capacitor, where the magnetic fields generated by pairs of symmetric current elements at point P cancel each other out, resulting in a zero net magnetic field.

elements (Fig. 4, right panel).

As mentioned, the magnetic field circulation on any closed curve *C* vanishes, however, when applying Ampère's law we find that there is a nonzero conduction current flowing through a surface $S_1$. This apparent contradiction can be understood if we observe that in non-stationary conditions Ampère's law is not compatible with the conservation of charge.

Consider a closed surface $S_0$ which encloses the inner conductor and a curve C dividing $S_0$ in two surfaces $S_1$ and $S_2$. Then, Ampère's law can be written in two different ways, either by considering the current through $S_1$

$$\oint \mathbf{B} \cdot \mathbf{dl} = \mu_0 \int_{S1} \mathbf{J} \cdot \hat{\mathbf{n}}_1 \mathrm{d}A \qquad (10)$$

or through $S_2$,

$$\oint \mathbf{B} \cdot d\mathbf{l} = \mu_0 \int_{S2} \mathbf{J} \cdot \hat{\mathbf{n}}_2 dA \qquad (11)$$

where $\hat{\mathbf{n}}_1$ is pointing outwards and $\hat{\mathbf{n}}_2$ inwards, so that both versors are direct in relation to the orientation of the curve *C*. If we subtract Eq. 10 to Eq. 11 we get

$$\mu_0 \int_{S1} \mathbf{J} \cdot \hat{\mathbf{n}}_1 dA - \mu_0 \int_{S2} \mathbf{J} \cdot \hat{\mathbf{n}}_2 dA = \mu_0 \oint_{S0} \mathbf{J} \cdot \hat{\mathbf{n}}_1 dA \qquad (12)$$

Note that since the magnetic field circulation is zero, both equations 10 and 11 and their subtraction should vanish.

$$\oint_{S0} \mathbf{J} \cdot \hat{\mathbf{n}}_1 dA = 0 \qquad (13)$$

This result, however, violates the principle of charge conservation, since the charge flow through $S_0$ implies a variation in the charge enclosed given by:

$$\oint_{S0} \mathbf{J} \cdot \mathbf{n}_1 dA + \frac{dq}{dt} = 0 \qquad (14)$$

where *q* is the charge inside $S_0$, or in this case the capacitor charge. According to Gauss's law, there exists an electric field flux on the surface $S_0$ given by

$$q = \oint_{S0} \varepsilon \mathbf{E} \cdot d\mathbf{A} \qquad (15)$$

Expressing Eq. 14 in terms **J** and **E** we find

$$0 = \mu_0 \left( \oint_{S0} \mathbf{J} \cdot d\mathbf{A} + \frac{d}{dt} \oint_{S0} \varepsilon \mathbf{E} \cdot d\mathbf{A} \right) \qquad (16)$$

in clear contradiction with Eq. 13.

The way to overcome this paradox is to rewrite Eqs. 10 and 11 including the new term

$$\oint_C \mathbf{B} \cdot d\mathbf{l} = \mu_0 \left( \int_{S1} \mathbf{J} \cdot d\mathbf{A} + \frac{d}{dt} \int_{S1} \varepsilon \mathbf{E} \cdot d\mathbf{A} \right) \qquad (17)$$

and

$$\oint_C \mathbf{B} \cdot d\mathbf{l} = \mu_0 \left( \int_{S2} \mathbf{J} \cdot d\mathbf{A} + \frac{d}{dt} \int_{S2} \varepsilon \mathbf{E} \cdot d\mathbf{A} \right) \qquad (18)$$

It is now easy to verify that both equations are not only equal to each other, but also that both are null, as we observed from the beginning. To do this we observe that the current density at a point at a distance *r* from the center is from the principle of charge conservation,

$$\mathbf{J} = -\frac{\dot{q}}{4\pi r^3}\mathbf{r} \quad (19)$$

and, by Gauss's law, the electric field at the same point is,

$$\mathbf{E} = \frac{q}{4\pi\varepsilon r^3}\mathbf{r} \quad (20)$$

Substituting Eqs. 19 and 20 in Eq. 17 we find that

$$\oint_C \mathbf{B} \cdot \mathbf{dl} = \mu_0 \int_{S1} \left(-\frac{\dot{q}}{4\pi r^3}\mathbf{r} + \frac{d}{dt}\frac{\varepsilon q}{4\pi\varepsilon r^3}\mathbf{r}\right) \cdot \mathbf{dA} \quad (21)$$

cancels out as we wanted to show.

In conclusion, from the point of view of the AM law, the magnetic field circulation on the curve $C$ is due to the presence of a displacement current which is exactly opposite to the conduction current. This example exhibits great versatility, since it can be used both to introduce the displacement current and the AM law and to emphasize, through a simple situation, that it implies the conservation of electric charge.

## 5. Fields in a capacitor and their relationships to Maxwell equations

Let us consider again a capacitor of parallel, circular plates of radius $R$, separated by a distance $d$ as shown in Fig. 5. The capacitor is charged by means of a power source, such that the voltage increases according to an arbitrary function $V(t)$. A standard textbook problem is to determine the magnetic field between the plates at a distance $r < R$ from the symmetry axis. To calculate it, let us apply the AM law using a circumference $C_1$ of radius $r$ in a plane parallel to the plates. Assuming the electric field between the plates is uniform in the longitudinal direction, the magnetic field circulation results

$$\oint_{C1} \mathbf{B} \cdot \mathbf{dl} = \mu_0 \varepsilon_0 \frac{d}{dt}\int \frac{V(t)}{d} dA \quad (22)$$

where $V(t)/d$ is the magnitude of the electric field between the plates. The magnitude of the magnetic field can be easily determined as

$$B = \frac{\mu_0 \varepsilon_0 r}{2d}\frac{dV}{dt} \quad (23)$$

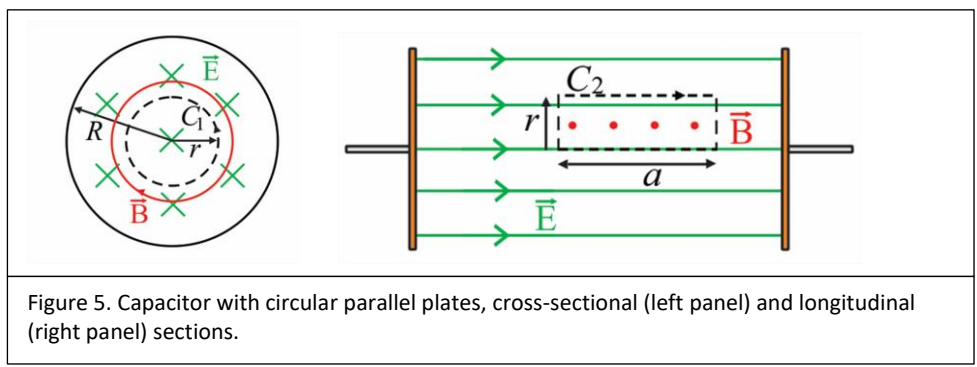

Figure 5. Capacitor with circular parallel plates, cross-sectional (left panel) and longitudinal (right panel) sections.

Since usually the AM law is dealt with in introductory courses after Faraday's law, it is interesting to check whether the magnetic field obtained satisfies this law. For this purpose, let us consider a closed curve $C_2$ depicted in Fig. 5. Using Eq. 23 Faraday's law can be expressed as

$$\oint_{C2} \mathbf{E} \cdot \mathbf{dl} = -\frac{d}{dt}\int \mathbf{B} \cdot \mathbf{dA} = -\frac{d}{dt}\int_0^r \frac{\mu_0 \varepsilon_0 r}{2d}\frac{dV}{dt} a\, dr = -\frac{\mu_0 \varepsilon_0 r^2}{4d}\frac{d^2 V}{dt^2} a \qquad (24)$$

Therefore, it predicts that the electric field circulation along $C_2$ is directly proportional to the second derivative of the voltage with respect to time. Now, as we assumed above, the electric field between the capacitor plates is uniform, its circulation along a closed curve must be zero. The only way this fact does not contradict Eq. 24, is that the second temporal derivative of the voltage vanishes, or equivalently, the voltage increases linearly with time as does the electric field between the plates. Under these conditions, since the displacement current is directly proportional to the time derivative of the voltage, Eq. 22, we conclude that a time-dependent uniform electric field can only exist if the displacement current, and hence the conduction current, are constant [24, 25]. This example reinforces the idea that except for artificial situations it is necessary to simultaneously solve Maxwell's equations to determine the fields [26, 27].

## 6. Current source and constant displacement current

A natural question that arises from the previous approach is how we could charge a capacitor so that the displacement current is constant. This is possible by charging the capacitor using a current source, i.e. a source that provides a constant current. Consider the circuit depicted in Fig. 6 where a current source $I_0$ is connected directly to a capacitor of capacitance $C$ initially discharged.

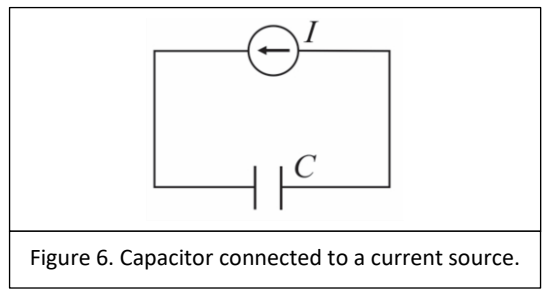

Figure 6. Capacitor connected to a current source.

The potential difference between the plates at time $t$ is $V(t) = Q(t)/C$ with $Q(t) = \int I_0 dt$. Assuming that initially the capacitor is discharged and integrating in time we obtain

$$V(t) = \frac{1}{C}\int I_0 dt = \frac{I_0}{C} t \qquad (25)$$

revealing that during the charging of a capacitor with a current source, the voltage increases linearly with time. Therefore, the electric field also varies proportionally, resulting in a displacement current constant between the capacitor plates.

This example highlights the underlying cause of the displacement current lies in the variation of the electric field, establishing a connection between these quantities independently of the specific characteristics of the conduction current. This analysis becomes particularly relevant from a didactic point of view, as it addresses a conceptual difficulty reported in the literature [11] related to the students' belief that there can only exist a displacement current when the conduction current changes with time. Although this idea may lead to correct conclusions in certain scenarios, it lacks generality. One possible cause of this type of reasoning is the association of displacement current with the standard textbook example of the capacitor charging across a voltage source where the current changes with time. Therefore, this analysis is beneficial in decoupling the displacement current from the characteristics of the conduction current, providing a more comprehensive and generalizable perspective.

## 7. Concluding remarks

In this paper, we have presented a set of situations that are plausible to be analyzed in electromagnetism courses. These situations allow us to clarify the Ampère-Maxwell law and the displacement current, as well as the nature of the relationships between the different terms in Maxwell's equations and the sources of the electromagnetic field. We hope that this work will encourage teachers of electromagnetics courses to broaden the range of situations studied in the context of the Ampère-Maxwell law and the displacement current. We believe that the examples analyzed offer a clear vision of their pedagogical value and can be used for the development of teaching materials, such as tutorials, to improve the understanding of electromagnetic theory.